\documentclass{PoS3}

\usepackage{amssymb, amsmath} 
\usepackage{cite}
\usepackage{float}
\newcommand{\GeV}{\mathrm{GeV}}

\catcode`,\active

\catcode`\,12

\title{{\footnotesize{DESY 18-119, DO-TH 18/16}}\\
Two-mass three-loop effects in deep-inelastic scattering}

\ShortTitle{
Two-mass three-loop effects in deep-inelastic scattering}

\author{ J.~Ablinger$^a$, J.~Bl\"umlein$^b$, A.~De Freitas$^b$, A.~Goedicke$^c$, C. 
Schneider$^a$, and \speaker{K.~Sch\"onwald}$^b$ \\
        \llap{$^a$}Research Institute for Symbolic Computation (RISC), Johannes Kepler 
        University, \\ 
                   Altenbergerstra\ss{}e 69, A--4040, Linz, Austria. \\
\llap{$^b$}Deutsches Elektronen--Synchrotron, DESY, \\
                   Platanenallee 6, D-15738 Zeuthen, Germany. \\
        \llap{$^c$}
Institut f\"ur Theoretische Teilchenphysik Campus S\"ud,
Karlsruher Institut f\"ur Technologie (KIT) D-76128 Karlsruhe, Germany}

\abstract{We report on recent results on the two-mass corrections for massive operator 
matrix elements at 2- and 3-loop orders in QCD. These corrections form the building blocks
of the variable flavor number scheme. Due to the similar values of the charm and bottom 
quark masses the two-mass corrections form an important contribution.}

\FullConference{ Loops and Legs in Quantum Field Theory - LL 2018,\\
		 29 April - 4 May 2018 \\
		 St. Goar, Germany }

\begin{document}
\section{Introduction}

\vspace*{1mm}
\noindent
The heavy flavor corrections to deep-inelastic scattering to 3-loop order are an important 
asset to describe the structure functions in the region of lower values of $x$ and to 
account for the scaling violations correctly, which are different in the massless and 
massive cases. They are required in precision measurements of the strong coupling constant
\cite{ALPHAS}, the determination of the heavy quark masses 
\cite{Alekhin:2012vu,Gizhko:2017fiu,Alekhin:2017kpj}, and the parton 
distribution functions \cite{Accardi:2016ndt,Alekhin:2017kpj,Alekhin:2018pai}.
In a long program, cf. \cite{Ablinger:2017ptf}, the different heavy flavor contributions 
to deep-inelastic 
scattering are calculated in the region $Q^2 \gg m^2$. For all quantities a series of 
Mellin moments has been obtained in \cite{Bierenbaum:2009mv}. The corrections in terms of the  
general Mellin variable $N$ in the single heavy mass case were calculated in the non-singlet, the 
pure singlet, the $gq$- and $gg$-cases 
\cite{Ablinger:2010ty,Ablinger:2014vwa,Behring:2015zaa,Behring:2015roa,Behring:2016hpa,
Ablinger:2014nga,Ablinger:2014lka,AGG}, as well as for all logarithmic corrections 
\cite{Behring:2014eya}. Likewise, we have also obtained all 3-loop anomalous dimensions 
$\gamma_{ij}$
contributing to the heavy flavor case in Refs.~\cite{Ablinger:2010ty,Ablinger:2014vwa,
Ablinger:2014nga,Ablinger:2017tan}, which are the terms $\propto T_F$. In case of 
$\gamma_{qq}^{\rm PS}$ and $\gamma_{qg}$ these are the complete anomalous dimensions.
All the 1st order factorizing contributions to the operator matrix element OME $A_{Qg}^{(3)}$, i.e. the 
contributions due to iterative integrals, have been computed \cite{Ablinger:2017ptf}. The 
calculation of these quantities required the development of 
several new computation techniques and algorithms, cf.~Ref.~\cite{Blumlein:2018cms} for 
a survey.

Already at 2-loop order, two-mass corrections contribute to the heavy flavor Wilson 
coefficients in the form of reducible terms, cf.~\cite{Ablinger:2017err}. However, for many 
years, they were not considered in the variable flavor number scheme 
(VFNS) \cite{Buza:1996wv}. 
At 3-loop order, genuine two-mass contributions appear \cite{Ablinger:2017err}. For them a 
number of moments has been calculated expanding in the mass ratio $\eta = 
m_c^2/m_b^2 \sim 1/10$, cf.~Ref.~\cite{Ablinger:2011pb,Ablinger:2012qj,Ablinger:2017err}. 
In the flavor 
non-singlet, transversity and $gq$-cases, the general $N$- and $x$-results have been 
calculated in Ref.~\cite{Ablinger:2017err}.

More recently, the complete 3-loop corrections in the flavor pure-singlet case 
\cite{Ablinger:2017xml} and for the massive OME $A_{gg}$ 
\cite{Ablinger:2018brx} have been computed. These results are discussed in 
Section~\ref{sec:2}. At 2-loop order all two-mass terms are known. In 
Section~\ref{sec:3} we discuss the variable flavor number scheme at 2-loop order, extended 
to the 2-mass case and describe the corresponding changes for the parton densities 
\cite{Blumlein:2018jfm}. They 
turn out to be of relevance for precision measurements at the LHC, comparing to the 
single mass case. Section~\ref{sec:4} contains the conclusions.
\section{\boldmath Three-Loop Corrections}
\label{sec:2}

\vspace*{1mm}
\noindent
Recently we computed the two-mass 3-loop corrections to the OMEs $A_{Qq}^{PS, \rm (3)}$ 
and $A_{gg,Q}^{\rm (3)}$ in Refs.~\cite{Ablinger:2017xml,Ablinger:2018brx}. Contrary to 
the 3-loop non-singlet cases and $A_{gq}$, the $\eta$-dependence does not factorize here.
The analytic results can be represented in terms of iterated integrals over general alphabets 
and specific integrals thereof. The iterated integrals are given by
\begin{eqnarray}
G_{f_0,f_1, ..., f_k}(x) = \int_0^x dy f_0(y) G_{f_1, ..., f_k}(y),~~~G_\emptyset = 1,
\label{eq:GG}
\end{eqnarray}
with letters $f_l(x)$, which may depend on the additional parameter $\eta$ and usually 
form root valued irrational functions. The package {\tt HarmonicSums} 
\cite{HSUM,Ablinger:2013cf,ABL18} 
allows 
the automated calculation of these integrals and the reduction of special constants which 
appear in this context.

It turns out that the calculation of the OME $A_{Qq}^{PS, \rm (3)}$ cannot be easily 
done in $N$ space. It is therefore computed in $x$-space and written in terms of a 
Mellin transform, separating a series of $N$-dependent pre-factors, which will be 
dealt 
with at a later stage. The contributing mass ratios are either ruled by $\eta$ or 
$1/\eta$. In the former case only one principle integration region is obtained, while in 
the latter case the regions $x \in \{[0,\eta_-],[\eta_-,\eta_+],[\eta_+,1]\}$ contribute, 
with $\eta_{\pm} = (1 \pm \sqrt{1-\eta})/2.$ Besides the letters of the usual harmonic 
polylogarithms (HPLs) \cite{Remiddi:1999ew}, two more letters 
\begin{eqnarray}
\sqrt{(4-x)x},~~~~~~\frac{\sqrt{1-4x}}{x}
\end{eqnarray}
contribute to the integrals of the Mellin transforms. Finally, one has to incorporate the 
$N$-dependent pre-factors by partial integration. From the global Mellin-transform one 
then obtains the two-mass contribution to the OME $A_{Qq}^{PS, \rm (3)}$. The structure  
of the constant term in $\varepsilon$ is given by
\begin{eqnarray}
a_{Qq}^{PS, \rm (3)}(N) &=& 
\int_0^1 dx \,\, x^{N-1} \biggl\{
K(\eta,x)
+\big(\theta(\eta_--x)+\theta(x-\eta_+)\big) x \, g_0(\eta,x)
\nonumber\\ &&
+\theta(\eta_+-x) \theta(x-\eta_-) \biggl[x \, f_0(\eta,x)
-{\int_{\eta_-}^x dy} \left(f_1(\eta,y)+\frac{y}{x} f_2(\eta,y)+\frac{x}{y}
f_3(\eta,y)\right)\biggr] 
\nonumber\\ &&
+\theta(\eta_--x) {\int_x^{\eta_-} dy} \left(g_1(\eta,y)+\frac{y}{x}
g_2(\eta,y)+\frac{x}{y} g_3(\eta,y)\right) \nonumber\\ &&
-\theta(x-\eta_+) {\int_{\eta_+}^x dy} \left(g_1(\eta,y)+\frac{y}{x}
g_2(\eta,y)+\frac{x}{y} g_3(\eta,y)\right) \nonumber\\ &&
+x \, h_0(\eta,x) + {\int_x^1 dy} \left(h_1(\eta,y)+\frac{y}{x} 
h_2(\eta,y)+\frac{x}{y}
h_3(\eta,y)\right) \nonumber\\ &&
+\theta(\eta_+-x) {\int_{\eta_-}^{\eta_+} dy} \left(f_1(\eta,y)+\frac{y}{\eta_+ x}
f_2(\eta,y)+\eta_+ \frac{x}{y} f_3(\eta,y)\right) \nonumber\\ &&
+{\int_{\eta_+}^1} dy \left(g_1(\eta,y)+\frac{y}{x} g_2(\eta,y)+\frac{x}{y}
g_3(\eta,y)\right)
\biggr\}.
\end{eqnarray}

\begin{figure}[H]
\begin{center}
\includegraphics[width=0.7\textwidth]{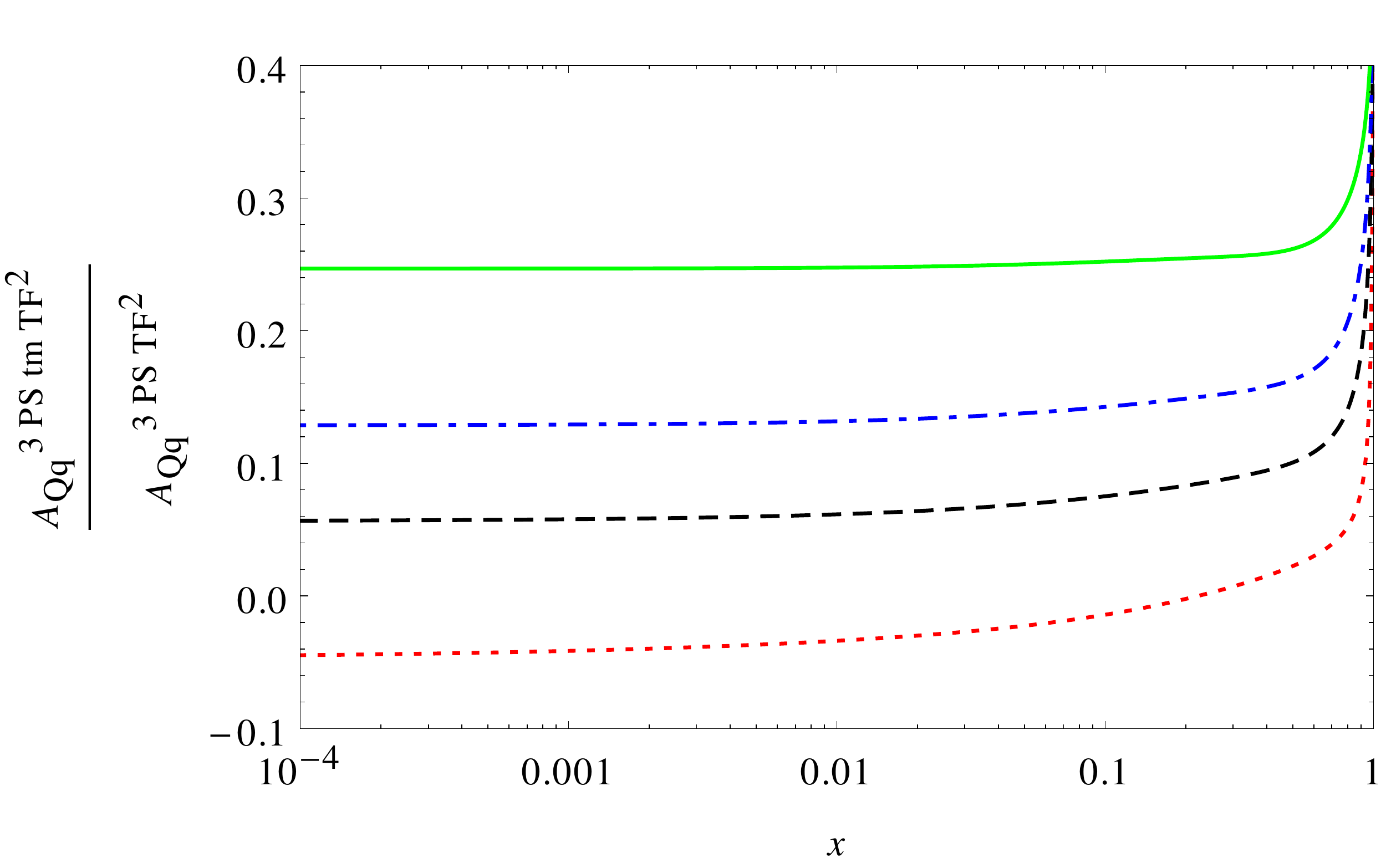}
\end{center}   
\caption{
The ratio of the 2-mass (tm) contributions to the massive OME $A_{Qq}^{{\sf PS},(3)}$ to 
all contributions to
$A_{Qq}^{{\sf PS},(3)}$
of $O(T_F^2)$ as a function of $x$ and $\mu^2$.
Dotted line (red): $\mu^2 = 30~\GeV^2$.
Dashed line (black): $\mu^2 = 50~\GeV^2$.
Dash-dotted line (blue): $\mu^2 = 100~\GeV^2$.
Full line (green): $\mu^2 = 1000~\GeV^2$. Here the on-shell heavy quark masses $m_c = 
1.59~\GeV$ and $m_b =
4.78~\GeV$
\cite{Alekhin:2012vu,Agashe:2014kda} have been used; from Ref.~\cite{Ablinger:2017xml}.}
\label{fig:FIG1}
\end{figure}
\begin{figure}[H]
\begin{center}
\includegraphics[width=0.7\textwidth]{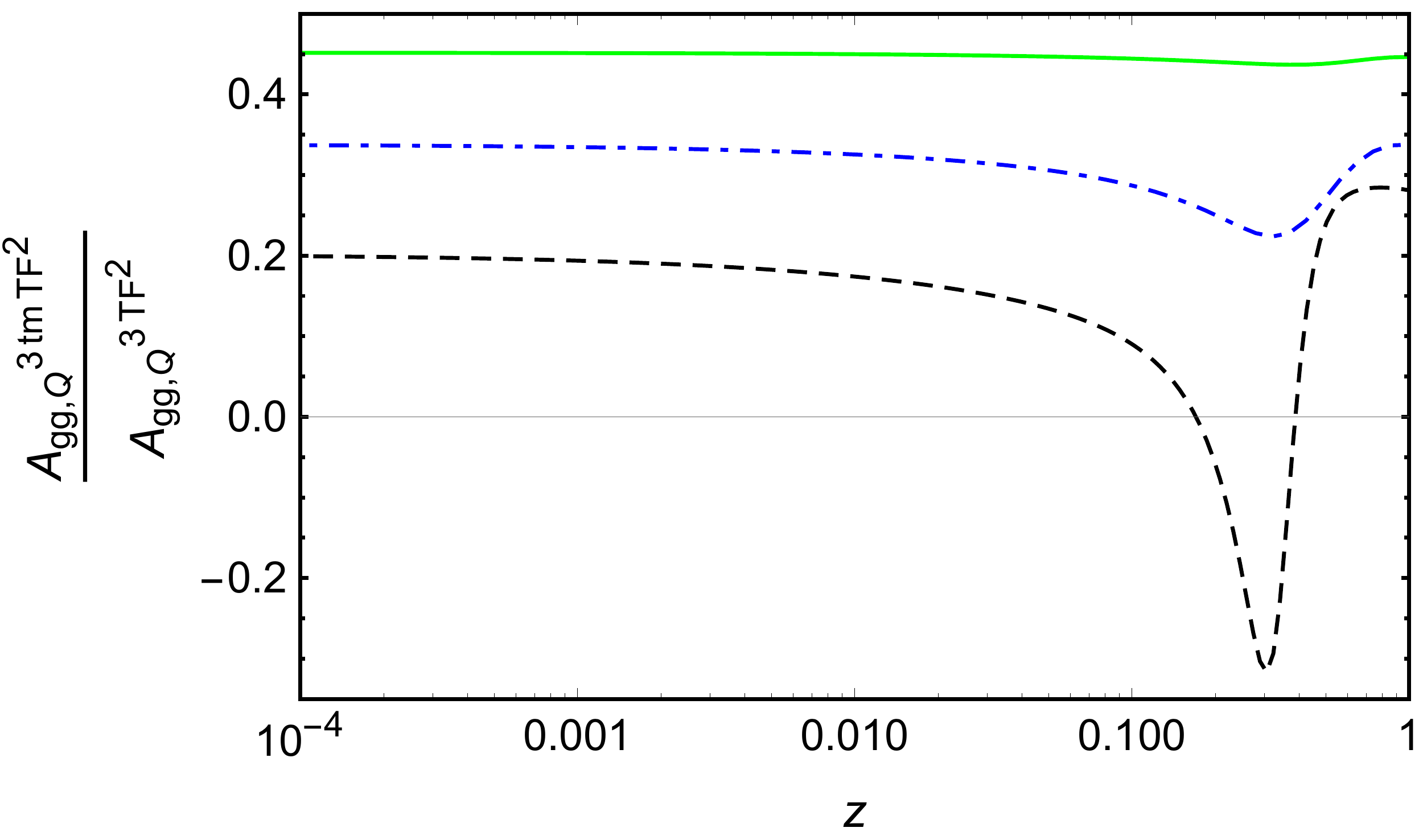}
\end{center}   
\caption{The ratio of the two-mass (tm) contributions to the massive OME 
$A_{gg,Q}^{(3)\rm{tm}}$ to all contributions
to $A_{gg,Q}^{(3)}$ of $O(T_F^2)$ as a function of $z$ and $\mu^2$.
Dashed line (black): $\mu^2 = 50~\GeV^2$.
Dash-dotted line (blue): $\mu^2 = 100~\GeV^2$.
Full line (green): $\mu^2 = 1000~\GeV^2$. Here the on-shell heavy quark masses $m_c = 
1.59~\GeV$ and $m_b =
4.78~\GeV$ \cite{Alekhin:2012vu,Agashe:2014kda} have been used; from 
Ref.~\cite{Ablinger:2018brx}.}
\label{fig:FIG2}
\end{figure}

The functions $K,f_i, g_i$ and $h_i$ are given in Ref.~\cite{Ablinger:2017xml}  and can be 
represented in terms of HPLs at more involved arguments for which the numerical 
representation is available \cite{Gehrmann:2001pz,Ablinger:2017tqs}. In 
Figure~\ref{fig:FIG2}  the ratio of the 3-loop two-mass corrections in the pure 
singlet case is compared to the complete $O(T_F^2 C_F (C_A))$. The ratio behaves about 
flat as a function of $x$, with some rise towards $x=1$ and grows with $\mu^2$ to typical 
values of $\sim 0.4$ and is therefore a significant contribution at this order.

In Ref.~\cite{Ablinger:2018brx} we have calculated the 3-loop two-mass contributions to 
the OME $A_{gg}$. Here the calculation can be either performed in $N$- or $x$-space.
In $N$-space it requires one analytic Mellin-Barnes integral aside of integrals which can 
be performed using simpler methods. We have chosen the direct calculation of all 
contributions, which leads to a large set of individual sum expressions. The largest 
diagram led to a representation of $\sim 100$MB. We use modern summation technologies
\cite{SUMMATION} encoded in the package {\tt Sigma} \cite{SIG1,SIG2}.
The sums are first crunched to a few 
master sums using the package {\tt SumProduction}. The latter sums are solved 
individually using {\tt EvaluatMultiSums} \cite{EMSSP} and limits to infinity are 
performed 
using built-in routines of {\tt HarmonicSums}. Finally the results are reduced to basic 
sums. In the case of the largest diagram, $\sim 78$ days were needed to perform the sums and 
$\sim 33$ days to reduce to the basis, i.e. to eliminate all relations between the sums.
The full summation of this OME took about five months.

Typical sums occurring are
\begin{eqnarray}
S_{\vec{k}}\left(\frac{\eta-1}{\eta},1,N\right),~~~\binom{2N}{N} \sum_{i=1}^N 
\frac{
4^i 
\left(\frac{\eta}{\eta-1}\right)^i
}{i \binom{2i}{i}}
S_{\vec{k}}\left(\frac{\eta-1}{\eta},1,i\right),~~~{\rm etc.}
 \end{eqnarray}
These are generalized harmonic sums at real weights,~cf.~e.g.~\cite{Ablinger:2013cf} and 
nested binomial 
sums over these objects, which generalize the class of sums having been dealt with in 
Ref.~\cite{Ablinger:2014bra} before. 

{\tt HarmonicSums} provides algorithms to perform the inverse Mellin transform to $x$ 
space. Here iterated integrals of the kind~(\ref{eq:GG}) occur. The letters of the 
corresponding alphabet are those of the usual HPLs and
\begin{eqnarray}
&&\sqrt{x(1-x)},~~
\frac{1}{x+\eta(1-x)},~~
\frac{1}{1-x(1-\eta)},~~
\frac{1}{\eta+x(1-\eta)},~~
\nonumber\\ && \hspace*{1cm}
\frac{\sqrt{x(1-x)}}{1-x(1-\eta)},~~
\frac{\sqrt{x(1-x)}}{x-\eta(1-x)},~~
\frac{\sqrt{x(1-x)}}{\eta+x(1-\eta)}.
 \end{eqnarray}
Again rational pre-factors in $N$ have to be absorbed like in the pure singlet case. This
leads to a one-dimensional integral representation over integrands out of HPLs with 
involved argument, well suited for a fast numerical evaluation. In 
Ref.~\cite{Ablinger:2018brx} detailed lists of integrals and special constants of the type 
(\ref{eq:GG}) emerging in the present context have also been given which can be used in 
similar calculations.
 
The relative effect of the 3-loop two-mass contributions to $A_{gg}$ in comparison to all 
$O(T_F^2 C_F$ $(C_A))$ terms is illustrated in dependence of $x$ and $\mu^2$. The ratio 
behaves flat in the small $x$ region, with some structure towards larger values of $x$. 
The ratio grows with $\mu^2$ and reaches values of $\sim 0.4$ at $\mu^2 = 1000 \GeV^2$.

\section{\boldmath Two-mass Corrections in the Variable Flavor Number Scheme}
\label{sec:3}

\vspace*{1mm}
\noindent
Since the mass ratio squared $\eta$ for charm and bottom is not a very small number, one 
cannot treat charm quarks as massless at the scale $\mu = m_b$. The decoupling in the 
variable flavor number scheme has therefore to account for the 2-mass effects from 
$O(\alpha_s^2)$ onward. In the usual VFNS, one decouples one heavy quark at a time, 
cf.~Ref.~\cite{Buza:1996wv}. Its generalization, cf.~Ref.~\cite{Blumlein:2018jfm}, 
accounts for the two-mass effects. The corresponding transition rules are:
\begin{eqnarray}
\label{eq:FI}
f_{{\sf NS},i}(N_F+2,\mu^2) &=& \Biggl\{1 + a_s^2(\mu^2) \left[
 A_{qq,Q}^{{\sf NS},(2,c)}
+A_{qq,Q}^{{\sf NS},(2,b)}\right] \Biggr\} f_{{\sf NS},i}(N_F,\mu^2),
\\
\label{eq:FIS}
\Sigma(N_F+2,\mu^2) &=& \Biggl\{1 + a_s^2(\mu^2)\Bigl[
 A_{qq,Q}^{{\sf NS},(2,c)} + A_{qq,Q}^{{\sf PS},(2,c)}
+A_{qq,Q}^{{\sf NS},(2,b)} + A_{qq,Q}^{{\sf PS},(2,b)} \Bigr] \Biggr\}
\Sigma(N_F,\mu^2)
\nonumber\\ &&
\hspace*{-3mm}
+ \Biggl\{a_s(\mu^2) \Bigl[
 A_{Qg}^{(1,c)}
+A_{Qg}^{(1,b)}\Bigr]
+a_s^2(\mu^2)
\Bigl[
 A_{Qg}^{(2,c)}
+A_{Qg}^{(2,b)} + A_{Qg}^{(2,cb)}\Bigr]\Biggr\} G(N_F,\mu^2),
\nonumber\\
\\
\label{eq:GNF2}
G(N_F+2,\mu^2) &=&
 \Biggl\{1 + a_s(\mu^2) \Bigl[
 A_{gg,Q}^{(1,c)}
+A_{gg,Q}^{(1,b)}\Bigr]
+a_s^2(\mu^2)
\Bigl[
 A_{gg,Q}^{(2,c)}
+A_{gg,Q}^{(2,b)} + A_{gg,Q}^{(2,cb)}\Bigr]\Biggr\} G(N_F,\mu^2)
\nonumber \\
&&
+a_s^2(\mu^2)\Bigl[
  A_{gq,Q}^{(2,c)}
+ A_{gq,Q}^{(2,b)} \Bigr] \Sigma(N_F,\mu^2), 
\\
&&
\label{eq:cc}
\lefteqn{\hspace*{-2.5cm}\Bigl[f_c+f_{\bar{c}}\Bigr](N_F+2,\mu^2) = a_s^2(\mu^2)
  A_{Qq}^{{\sf PS},(2,c)}
 \Sigma(N_F,\mu^2)
}
\nonumber\\
&&~~~~~~
+ \Biggl\{a_s(\mu^2)
A_{Qg}^{(1,c)}
+a_s^2(\mu^2)
\Bigl[
 A_{Qg}^{(2,c)}
+ \frac{1}{2} A_{Qg}^{(2,cb)}\Bigr]\Biggr\} G(N_F,\mu^2),
\\
&&
\label{eq:ma2VNS}
\lefteqn{\hspace*{-2.5cm}\Bigl[f_b+f_{\bar{b}}\Bigr](N_F+2,\mu^2) = a_s^2(\mu^2)
  A_{Qq}^{{\sf PS},(2,b)}
 \Sigma(N_F,\mu^2)
}
\nonumber\\
&&~~~~~~
+ \Biggl\{a_s(\mu^2)
A_{Qg}^{(1,b)}
+a_s^2(\mu^2)
\Bigl[
 A_{Qg}^{(2,b)}
+ \frac{1}{2} A_{Qg}^{(2,cb)}\Bigr]\Biggr\} G(N_F,\mu^2)~,
\end{eqnarray}
where $a_s = \alpha_s/(4\pi)$.
\begin{figure}[H]
\begin{center}
\includegraphics[width=0.48\textwidth]{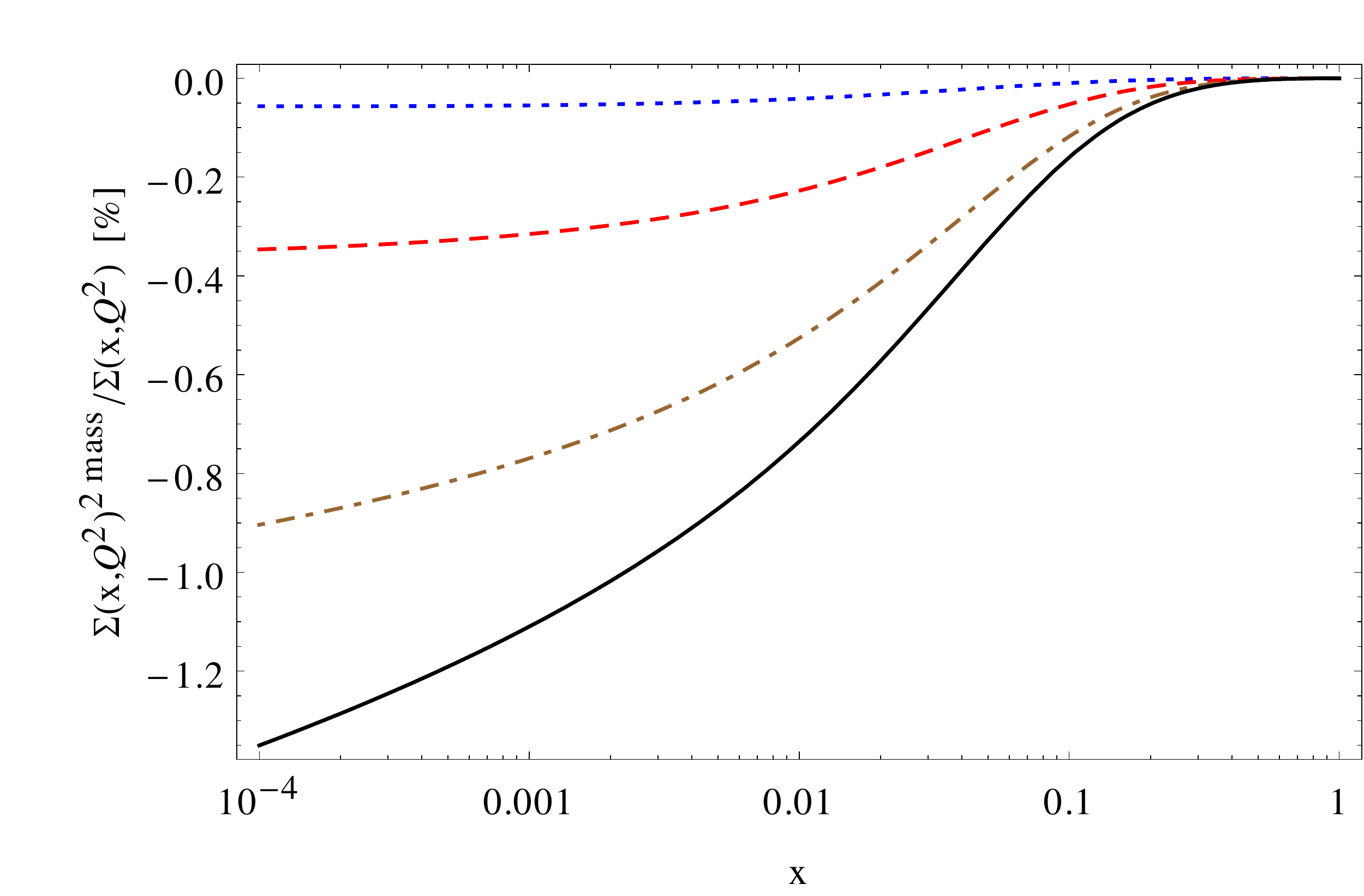}
\includegraphics[width=0.48\textwidth]{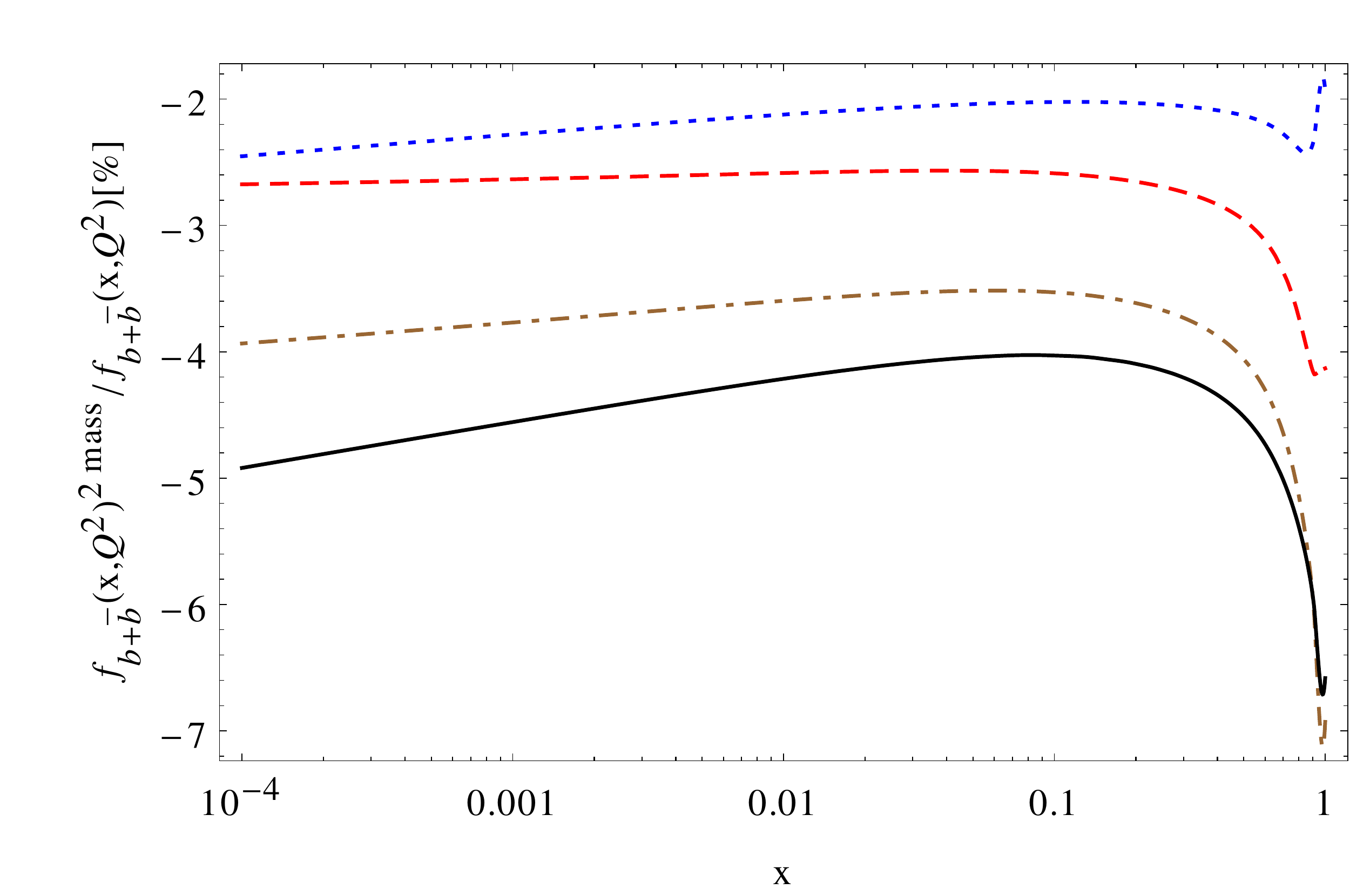}
\end{center}   
\caption{The ratio of the two-mass contribution to the
singlet distribution and the complete singlet distribution at $O(a_s^2)$, 
Eq.~(\ref{eq:FIS}), in \%, as a
function of $x$ and $Q^2$,
using the parton distribution functions
\cite{Alekhin:2017kpj} and $m_c =1.59~\GeV$ \cite{Alekhin:2012vu}, $m_b = 4.78~\GeV$ 
\cite{Agashe:2014kda}.
Dotted line: $Q^2 = 30~\GeV^2$;
dashed line: $Q^2 = 100~\GeV^2$;
dash-dotted line: $Q^2 = 1000~\GeV^2$;
full line: $Q^2 = 10000~\GeV^2$; from Ref.~\cite{Blumlein:2018jfm}.}
\label{fig:FIG3}
\end{figure}
Comparing the two-mass contributions to the different parton distribution functions with 
the complete result one finds $O(1\%)$ effects e.g. for the flavor singlet distribution.
For the bottom quark distribution the effect can reach $O(5\%)$ at typical scales $Q^2 \in 
[30, 10000] \GeV^2$ at the LHC, see Figure~\ref{fig:FIG3}.

At 3-loop order still the two-mass corrections to the OME $A_{Qg}^{(3)}$ have to be 
calculated in order to describe the generalized VFNS. This is work in progress and currently 
$O(500)$ even moments have been calculated, expanding in the mass ratio $\eta$ to the 
5th order.
\section{Conclusions}
\label{sec:4}

\vspace*{1mm}
\noindent
The 2-mass contributions to the massive OMEs in deep-inelastic scattering have a 
significant numerical effect on a series of parton distributions in precision data 
analysis. Already at 2-loop order the bottom quark distribution at the LHC receives
about 5\% corrections, while those for other PDFs are smaller, but not negligible at the 
1\% level. We have by now calculated all two-mass contributions at 3-loop order, but those 
for the massive OME $A_{Qg}$. In the latter case, iterative integrals will not be 
sufficient to represent this quantity and at least iterated integrals over elliptic 
integrals are contributing as well, cf.~\cite{Ablinger:2017bjx,Blumlein:2018aeq}. In an 
ongoing study we investigate first a large number of Mellin 
moments expanding in the mass ratio $\eta$ working towards a two-mass generalization of 
the VFNS also at 3-loop order. Along with the present calculations, again new powerful 
analytic integration techniques have been designed, which can be used in other 2- and 
3-loop calculations facing two-scale problems.

\vspace*{3mm}
\noindent
{\bf Acknowledgment.}~
This work was supported in part by the Austrian Science Fund (FWF) grant SFB F50 
(F5009-N15).

 

\begin{thebibliography}{99}
%
\bibitem{ALPHAS}
  S.~Bethke {\it et al.},
  {\it Workshop on Precision Measurements of alphas},
  arXiv:1110.0016 [hep-ph];\\
  S.~Moch, S.~Weinzierl {\it et al.},
  {\it High precision fundamental constants at the TeV scale},
  arXiv:1405.4781 [hep-ph];\\
  S.~Alekhin, J.~Bl\"umlein and S.O.~Moch,
  Mod.\ Phys.\ Lett.\ A {\bf 31} (2016) no.25,  1630023.
%
\bibitem{Alekhin:2012vu}
  S.~Alekhin, J.~Bl\"umlein, K.~Daum, K.~Lipka and S.~Moch,
  Phys.\ Lett.\ B {\bf 720} (2013) 172
  [arXiv:1212.2355 [hep-ph]].
%
\bibitem{Gizhko:2017fiu}
  A.~Gizhko {\it et al.},
  Phys.\ Lett.\ B {\bf 775} (2017) 233
  [arXiv:1705.08863 [hep-ph]].
%
\bibitem{Alekhin:2017kpj}
  S.~Alekhin, J.~Bl\"umlein, S.~Moch and R.~Placakyte,
  Phys.\ Rev.\ D {\bf 96} (2017) no.1,  014011
  [arXiv:1701.05838 [hep-ph]].
%
\bibitem{Accardi:2016ndt}
  A.~Accardi {\it et al.},
  Eur.\ Phys.\ J.\ C {\bf 76} (2016) no.8,  471
  [arXiv:1603.08906 [hep-ph]].
%
\bibitem{Alekhin:2018pai}
  S.~Alekhin, J.~Bl\"umlein and S.~Moch,
  Eur.\ Phys.\ J.\ C {\bf 78} (2018) no.6,  477
  [arXiv:1803.07537 [hep-ph]].
%
\bibitem{Ablinger:2017ptf}
  J.~Bl\"umlein, J.~Ablinger, A.~Behring, A.~De Freitas, A.~von Manteuffel, C.~Schneider and 
  C.~Schneider,
  PoS (QCDEV2017)  031
  [arXiv:1711.07957 [hep-ph]].
%
\bibitem{Bierenbaum:2009mv}
  I.~Bierenbaum, J.~Bl\"umlein and S.~Klein,
  Nucl.\ Phys.\ B {\bf 820} (2009) 417
  [arXiv:0904.3563 [hep-ph]];\\
  J.~Bl\"umlein, S.~Klein and B.~T\"odtli,
  Phys.\ Rev.\ D {\bf 80} (2009) 094010
  [arXiv:0909.1547 [hep-ph]].
%
\bibitem{Ablinger:2010ty}
  J.~Ablinger, J.~Bl\"umlein, S.~Klein, C.~Schneider and F.~Wi\ss{}brock,
  Nucl.\ Phys.\ B {\bf 844} (2011) 26
  [arXiv:1008.3347 [hep-ph]].
%
\bibitem{Ablinger:2014vwa}
J.~Ablinger, A.~Behring, J.~Bl\"umlein, A.~De Freitas, A.~Hasselhuhn, A.~von Manteuffel, 
M.~Round, C.~Schneider, and F.~Wi\ss{}brock, 
  Nucl.\ Phys.\ B {\bf 886} (2014) 733
  [arXiv:1406.4654 [hep-ph]].
%
\bibitem{Behring:2015zaa}
  A.~Behring, J.~Bl\"umlein, A.~De Freitas, A.~von Manteuffel and C.~Schneider,
  Nucl.\ Phys.\ B {\bf 897} (2015) 612
  [arXiv:1504.08217 [hep-ph]].
%
\bibitem{Behring:2015roa}
  A.~Behring, J.~Bl\"umlein, A.~De Freitas, A.~Hasselhuhn, A.~von Manteuffel and C.~Schneider,
  Phys.\ Rev.\ D {\bf 92} (2015) no.11,  114005
  [arXiv:1508.01449 [hep-ph]].
%
\bibitem{Behring:2016hpa}
  A.~Behring, J.~Bl\"umlein, G.~Falcioni, A.~De Freitas, A.~von Manteuffel and C.~Schneider,
  Phys.\ Rev.\ D {\bf 94} (2016) no.11,  114006
  [arXiv:1609.06255 [hep-ph]].
%
\bibitem{Ablinger:2014nga}
  J.~Ablinger, A.~Behring, J.~Bl\"umlein, A.~De Freitas, A.~von Manteuffel and C.~Schneider,
  Nucl.\ Phys.\ B {\bf 890} (2014) 48
  [arXiv:1409.1135 [hep-ph]].
%
\bibitem{Ablinger:2014lka}
  J.~Ablinger, J.~Bl\"umlein, A.~De Freitas, A.~Hasselhuhn, A.~von Manteuffel, M.~Round, 
  C.~Schneider and F.~Wi\ss{}brock,
  Nucl.\ Phys.\ B {\bf 882} (2014) 263
  [arXiv:1402.0359 [hep-ph]].
%
\bibitem{AGG}
J.~Ablinger, A.~Behring, J.~Bl\"umlein, A.~De Freitas, A.~von Manteuffel, and 
C.~Schneider, DESY 15--112.
%
\bibitem{Behring:2014eya}
  A.~Behring, I.~Bierenbaum, J.~Bl\"umlein, A.~De Freitas, S.~Klein and F.~Wi\ss{}brock,
  Eur.\ Phys.\ J.\ C {\bf 74} (2014) no.9,  3033
  [arXiv:1403.6356 [hep-ph]].
%
\bibitem{Ablinger:2017tan}
  J.~Ablinger, A.~Behring, J.~Bl\"umlein, A.~De Freitas, A.~von Manteuffel and C.~Schneider,
  Nucl.\ Phys.\ B {\bf 922} (2017) 1
  [arXiv:1705.01508 [hep-ph]].
%
\bibitem{Blumlein:2018cms}
  J.~Bl\"umlein and C.~Schneider,
  Int.\ J.\ Mod.\ Phys.\ A {\bf 33} (2018) no.17,  1830015.
%
\bibitem{Ablinger:2017err}
  J.~Ablinger, J.~Bl\"umlein, A.~De Freitas, A.~Hasselhuhn, C.~Schneider and 
  F.~Wi\ss{}brock,
  Nucl.\ Phys.\ B {\bf 921} (2017) 585
  [arXiv:1705.07030 [hep-ph]].
%
\bibitem{Buza:1996wv}
  M.~Buza, Y.~Matiounine, J.~Smith and W.~L.~van Neerven,
  Eur.\ Phys.\ J.\ C {\bf 1} (1998) 301
  [hep-ph/9612398].
%
\bibitem{Ablinger:2011pb}
  J.~Ablinger, J.~Bl\"umlein, S.~Klein, C.~Schneider and F.~Wi\ss{}brock,
  arXiv:1106.5937 [hep-ph].
%
\bibitem{Ablinger:2012qj}
  J.~Ablinger, J.~Bl\"umlein, A.~Hasselhuhn, S.~Klein, C.~Schneider and F.~Wi\ss{}brock,
  \newline PoS (RADCOR2011)  031
  [arXiv:1202.2700 [hep-ph]].
%
\bibitem{Ablinger:2017xml}
  J.~Ablinger, J.~Bl\"umlein, A.~De Freitas, C.~Schneider and K.~Sch\"onwald,
  Nucl.\ Phys.\ B {\bf 927} (2018) 339
  [arXiv:1711.06717 [hep-ph]].
%
\bibitem{Ablinger:2018brx}
  J.~Ablinger, J.~Bl\"umlein, A.~De Freitas, A.~Goedicke, C.~Schneider and K.~Sch\"onwald,
  Nucl.\ Phys.\ B {\bf 932} (2018) 129
  [arXiv:1804.02226 [hep-ph]].
%
\bibitem{Blumlein:2018jfm}
  J.~Bl\"umlein, A.~De Freitas, C.~Schneider and K.~Sch\"onwald,
  Phys.\ Lett.\ B {\bf 782} (2018) 362
  [arXiv:1804.03129 [hep-ph]].
%
\bibitem{HSUM}
  J.~Ablinger,
  PoS {(LL2014)} 019;
  {\it Computer Algebra Algorithms for Special Functions in Particle Physics}, Ph.D. 
Thesis, J. Kepler University
Linz, 2012,
  arXiv:1305.0687 [math-ph];\\
  {\it A Computer Algebra Toolbox for Harmonic Sums Related to Particle Physics}, Diploma 
Thesis, J. Kepler University Linz, 2009,
  arXiv:1011.1176 [math-ph];\\
  J.~Ablinger, J.~Bl\"umlein and C.~Schneider,
  J.\ Math.\ Phys.\  {\bf 52} (2011) 102301
  [arXiv:1105.6063 [math-ph]];\\
  J.~Ablinger,
  PoS (RADCOR2017) 001.
%
\bibitem{Ablinger:2013cf}
  J.~Ablinger, J.~Bl\"umlein and C.~Schneider,
  J.\ Math.\ Phys.\  {\bf 54} (2013) 082301
  [arXiv:1302.0378 [math-ph]];\\
%
\bibitem{ABL18}
J.~Ablinger, {\it An improved Algorithm to compute Inverse Mellin Transforms of Nested 
Binomial Sums}, PoS (LL2018) 063.
%
\bibitem{Remiddi:1999ew}
  E.~Remiddi and J.A.M.~Vermaseren,
  Int.\ J.\ Mod.\ Phys.\ A {\bf 15} (2000) 725
  [hep-ph/9905237].
%
\bibitem{Agashe:2014kda}
  K.~A.~Olive {\it et al.} [Particle Data Group],
  Chin.\ Phys.\ C {\bf 38} (2014) 090001.
%
\bibitem{Gehrmann:2001pz}
  T.~Gehrmann and E.~Remiddi,
  Comput.\ Phys.\ Commun.\  {\bf 141} (2001) 296
  [hep-ph/0107173].
%
\bibitem{Ablinger:2017tqs}
  J.~Ablinger, J.~Bl\"umlein, M.~Round and C.~Schneider,
  PoS (RADCOR2017) 010
  [arXiv:1712.08541 [hep-th]].
%
\bibitem{SUMMATION}
 M.~Karr, {J.~ACM} {\bf 28} (1981) 305;\\
M.~Bronstein, { J.~Symbolic Comput.} {\bf 29} (2000), no.~6 841;\\
 C.~Schneider, {\it Symbolic Summation in Difference Fields\/}, Ph.D. Thesis
  RISC, Johannes Kepler University, Linz technical report 01--17 (2001);
An. Univ. Timisoara Ser. Mat.-Inform. {\bf 42} (2004) 163;
{J. Differ. Equations Appl.} {\bf 11} (2005) 799;
Appl. Algebra Engrg. Comm. Comput. {\bf16} (2005) 1;
{J. Algebra Appl.} {\bf 6} (2007) 415.
{\it Motives, Quantum Field Theory, and Pseudodifferential
  Operators}, {\it Clay Mathematics Proceedings} Vol.~{\bf{12}}, eds. A.~Carey,
  D.~Ellwood, S.~Paycha and S.~Rosenberg, (Amer. Math. Soc) (2010), 285,
  [arXiv:0904.2323];
{Ann. Comb.} {\bf 14} (2010) 533 [arXiv:0808.2596];\\
 C.~Schneider, in: {\it Computer Algebra and Polynomials, Applications of
  Algebra and Number Theory}, J.~Gutierrez, J.~Schicho, M.~Weimann (ed.),
  Lecture Notes in Computer Science (LNCS) 8942 (2015), 157;
  [arXiv:1307.7887 [cs.SC]];
{J. Symbolic Comput.} {\bf 43} (2008) 611
  [arXiv:0808.2543];
J. Symb. Comput. {\bf 72} (2016) 82 [arXiv:1408.2776
  [cs.SC]];\\
J. Symb. Comput. {\bf 80} (2017) 616 [arXiv:1603.04285
  [cs.SC]].
%
\bibitem{SIG1}
C.~Schneider, {S\'em.~Lothar. Combin.\/} {\bf 56} (2007) 1, 
 article B56b.
%
\bibitem{SIG2}
C.~Schneider, {{\it Computer Algebra in Quantum Field Theory: Integration,
  Summation and Special Functions}\/} Texts and Monographs in Symbolic
  Computation eds. C.~Schneider and J.~Bl\"umlein  (Springer, Wien, 2013) 325, 
  arXiv:1304.4134 [cs.SC].
%
\bibitem{EMSSP}
  J.~Ablinger, J.~Bl\"umlein, S.~Klein and C.~Schneider,
  Nucl.\ Phys.\ Proc.\ Suppl.\  {\bf 205-206} (2010) 110
  [arXiv:1006.4797 [math-ph]];\\
  J.~Bl\"umlein, A.~Hasselhuhn and C.~Schneider,
  PoS (RADCOR 2011) 032
  [arXiv:1202.4303 [math-ph]];\\
  C.~Schneider,
  J.\ Phys.\ Conf.\ Ser.\  {\bf 523} (2014) 012037
  [arXiv:1310.0160 [cs.SC]].
%
\bibitem{Ablinger:2014bra}
  J.~Ablinger, J.~Bl\"umlein, C.G.~Raab and C.~Schneider,
  J.\ Math.\ Phys.\  {\bf 55} (2014) 112301
  [arXiv:1407.1822 [hep-th]].
%
\bibitem{Ablinger:2017bjx}
  J.~Ablinger, J.~Bl\"umlein, A.~De Freitas, M.~van Hoeij, E.~Imamoglu, C.~G.~Raab, C.-S.~Radu and C.~Schneider,
  J.\ Math.\ Phys.\  {\bf 59} (2018) no.6,  062305
  [arXiv:1706.01299 [hep-th]].
%
\bibitem{Blumlein:2018aeq}
  J.~Bl\"umlein, A.~De Freitas, M.~van Hoeij, E.~Imamoglu, P.~Marquard and C.~Schneider,
  PoS (LL2018) 017 [arXiv:1807.05287 [hep-ph]].
\end{thebibliography}
\end{document}